\documentclass[aps,prb,
amsmath,amssymb,
 preprint, groupedaddress, superscriptaddress,
 showpacs]{revtex4-1}

\usepackage{bm}
\usepackage{graphicx}
\usepackage{gensymb}
\usepackage{amsmath}
\usepackage{amssymb}
\usepackage{color}
\usepackage{ulem}
\usepackage{hyperref}
\hypersetup{
colorlinks = true,
urlcolor = blue,
linkcolor=blue,
citecolor = blue      
}

\begin{document}

\title{Observation of giant band splitting in altermagnetic MnTe}%

\author{T. Osumi}
\affiliation{Department of Physics, Graduate School of Science, Tohoku University, Sendai 980-8578, Japan}

\author{S. Souma}
\thanks{Corresponding authors:\\
s.souma@arpes.phys.tohoku.ac.jp\\
t-sato@arpes.phys.tohoku.ac.jp}
\affiliation{Center for Science and Innovation in Spintronics (CSIS), Tohoku University, Sendai 980-8577, Japan}
\affiliation{Advanced Institute for Materials Research (WPI-AIMR), Tohoku University, Sendai 980-8577, Japan}

\author{T. Aoyama}
\affiliation{Department of Physics, Graduate School of Science, Tohoku University, Sendai 980-8578, Japan}

\author{K. Yamauchi}
\affiliation{Center for Spintronics Research Network (CSRN), Osaka University, Toyonaka, Osaka 560–8531, Japan}

\author{A. Honma}
\affiliation{Department of Physics, Graduate School of Science, Tohoku University, Sendai 980-8578, Japan}

\author{K. Nakayama}
\affiliation{Department of Physics, Graduate School of Science, Tohoku University, Sendai 980-8578, Japan}

\author{T. Takahashi}
\affiliation{Department of Physics, Graduate School of Science, Tohoku University, Sendai 980-8578, Japan}

\author{K. Ohgushi}
\affiliation{Department of Physics, Graduate School of Science, Tohoku University, Sendai 980-8578, Japan}

\author{T. Sato}
\thanks{Corresponding authors:\\
s.souma@arpes.phys.tohoku.ac.jp\\
t-sato@arpes.phys.tohoku.ac.jp}
\affiliation{Department of Physics, Graduate School of Science, Tohoku University, Sendai 980-8578, Japan}
\affiliation{Center for Science and Innovation in Spintronics (CSIS), Tohoku University, Sendai 980-8577, Japan}
\affiliation{Advanced Institute for Materials Research (WPI-AIMR), Tohoku University, Sendai 980-8577, Japan}
\affiliation{International Center for Synchrotron Radiation Innovation Smart (SRIS), Tohoku University, Sendai 980-8577, Japan}
\affiliation{Mathematical Science Center for Co-creative Society (MathCCS), Tohoku University, Sendai 980-8577, Japan}

\date{\today}

\begin{abstract}
We performed angle-resolved photoemission spectroscopy (ARPES) on hexagonal MnTe, a candidate for an altermagnet with a high critical temperature ($T_\mathrm{N}=307 \textrm{ K}$). By utilizing photon-energy-tunable ARPES in combination with first-principles calculations, we found that the band structure in the antiferromagnetic phase exhibits a strongly anisotropic band-splitting associated with the time-reversal-symmetry breaking, providing the first direct experimental evidence for the altermagnetic band-splitting. The magnitude of the splitting reaches 0.8 eV at non-high-symmetry momentum points, which is much larger than the spin-orbit gap of $\sim$0.3 eV along the $\Gamma\textrm{K}$ high-symmetry cut. The present result paves the pathway toward realizing exotic physical properties associated with the altermagnetic spin-splitting. 
\end{abstract}


\maketitle

\clearpage
\section{INTRODUCTION}
A new type of magnetism called ``altermagnetism'' is attracting a great deal of attention because it is regarded as the third type of magnetism neither categorized into conventional ferromagnetism nor antiferromagnetism \cite{SmejkalPRX2022_1, SmejkalPRX2022_2}. In ferromagnets, the energy bands are spin split due to the breaking of time-reversal symmetry (TRS). In colinear antiferromagnets, the energy bands are generally spin degenerate because of the antiparallel configuration of magnetic moment and the zero net magnetization. Intriguingly, altermagnets are different from these two magnets; despite the zero magnetization as in antiferromagnets, the bands in altermagnets are spin split due to the TRS breaking as in ferromagnets. This band splitting is guaranteed by the existence of opposite-spin sublattices connected by the crystal-rotation symmetries \cite{SmejkalPRX2022_1, SmejkalPRX2022_2, HayamiJPSJ2019, YuanPRB2020, HYMaNC2021}. In contrast to ferromagnets where the band splitting is typically isotropic in the momentum ({\it{k}}) space, altermagnets exhibit an anisotropic band-splitting. This is because the sign of band splitting must be reversed between the positive and negative {\it{k}}’s to meet the requirement from the zero net magnetization, leading to band splitting with nodes on high-symmetry lines in the bulk Brillouin zone (BZ) under the absence of the spin-orbit coupling (SOC). Although an anisotropic band-splitting appears in nonmagnets as a consequence of space-inversion symmetry breaking and SOC \cite{Rashba1960}, the altermagnetic band-splitting is  distinct from it, because the splitting appears even without SOC \cite{SmejkalPRX2022_1, SmejkalPRX2022_2, HayamiJPSJ2019, YuanPRB2020, HYMaNC2021, HayamiPRB2020, YuanPRM2021}. Besides the fundamental interest in the unconventional band splitting, the altermagnetic band-splitting is predicted to host a variety of exotic quantum phenomena such as the anomalous Hall effect, spin current  generation, spin splitter torque, magnetic circular dichroism, crystal Nernst effect, and piezomagnetic effect \cite{SmejkalPRX2022_1, SmejkalPRX2022_2, HYMaNC2021, NakaNC2019, SmejkalSciAdv2020, NakaPRB2021, MazinPNAS2021, ThaoPRB2023, DFShaoPRL2023, Rooj2024, MazinPRB2023, ZhouPRL2024}. Some of them were recently verified experimentally \cite{XChenNM2021, BaiPRL2022, KarubePRL2022, BoseNatElec2022, BaiPRL2022, GonzalezPRL2023, BaiPRL2023, HarikiArXiv2023, AoyamaArXiv2023, FedchenkoArXiv2023}. However, the direct experimental evidence for the altermagnetic band-splitting, which would serve as a base to understand the exotic phenomena of altermagnets, is still missing. Such a lifting of spin degeneracy in the energy bands was unexpected for conventional antiferromagnets and indeed has never been observed to date. It is thus urgently required to experimentally verify the altermagnetic band-splitting to establish this new class of magnet and further advance the exciting physics of unconventional magnets.

Here we focus on hexagonal MnTe which is suited to observe the possible large spin splitting \cite{SmejkalPRX2022_1, Rooj2024, MazinPRB2023} due to its  high N{\'{e}}el temperature ($T_\mathrm{N}=307 \textrm{ K}$) \cite{Squire1939, Komatsubara1963, Szuszkiewicz2005, Efrem2005}. MnTe crystallizes in the NiAs-type structure (space group: $P6_3/mmc$) consisting of alternately stacked Mn and Te planes characterized by a simple hexagonal Mn lattice and a hexagonal closed packed Te lattice with the AB stacking [for crystal structure and bulk BZ, see Figs. 1(a) and 1(b), respectively] \cite{JuzaZ1956}. In the antiferromagnetic (AF) phase, the Mn$^{2+}$ ions placed at the center of the Te octahedra are ferromagnetically aligned within the same Mn plane with the magnetic moment lying in the (0001) plane, whereas the Mn ions between adjacent Mn planes are antiferromagnetically coupled \cite{Komatsubara1963, Szuszkiewicz2005,KriegnerPRB2017}. Intriguingly, opposite-spin sublattices essential for the altermagnetism are realized in the Mn sites because the octahedral coordination is inequivalent between the AB- and BA-stacked Mn sites. This situation does not allow spin reversal with a single symmetry operation and leads to the breaking of global TRS. The DFT (density-functional-theory) calculations predict a giant spin splitting \cite{SmejkalPRX2022_1, Rooj2024, MazinPRB2023, PauloPRB2023}, whereas such a splitting has yet to be clarified. 

 In this article, we report an ARPES study of MnTe bulk single crystal and show a direct evidence for the altermagnetic band-splitting. This was enabled by the utilization of photon-energy tunable micro-focused ARPES in combination with DFT calculations. We found that the magnitude of altermagnetic band-splitting strongly depends on the $k$ location in the 3D BZ, and is also affected by the SOC. We discuss implications of the present results in relation to other experiments on altermagnets.
 
 \section{EXPERIMENTS}
Single crystals of MnTe were grown by the chemical vapor transport method with I$_2$ as an agent gas. Obtained single crystals were characterized by x-ray diffraction, magnetization, and electrical resistivity measurements. Details of the sample preparation and characterization were described elsewhere \cite{AoyamaArXiv2023} (see also Appendix\ \hyperref[apxA]{A}). ARPES measurements were performed with micro-focused vacuum-ultraviolet (VUV) synchrotron light at BL-28A in KEK-PF \cite{KitamuraRSI}. We used linearly polarized light of 60--170 eV. The energy resolution was set to be 10--20 meV. Samples were cleaved {\it in situ} along the (0001) plane of hexagonal crystal in an ultrahigh vacuum of $1\times10^{-10} \textrm{ Torr}$. The crystal orientation was determined by the x-ray Laue backscattering measurement consistent with the (0001) cleaved plane. Since MnTe single crystal is hard to cleave, we tried to cleave several times in an ultrahigh vacuum (UHV), occasionally obtained a small flat area of the crystal with a few tens $\mu\mathrm{m}$ square and then focused the micro photon beam on it. Temperature during ARPES measurements was set at $T=30\text{--}310 \textrm{ K}$, which covers $T_\mathrm{N}$ (= 307 K). The Fermi level ($E_\mathrm{F}$) of samples was referenced to that of a gold film electrically in contact with the sample holder. First-principles band-structure calculations were carried out by using a projector augmented wave method implemented in Vienna Ab initio Simulation Package (VASP) code \cite{Kresse1996} with generalized gradient approximation (GGA) pseudopotential. The lattice constants were fixed to the experimental values ($a=b=4.158 \textrm{ \AA}$ and $c=6.726 \textrm{ \AA}$) \cite{LiChemMat2022}. The total energy was calculated self-consistently with the tetrahedron sampling of $12\times12\times8$ $k$-point mesh taking into account SOC.

 \section{RESULTS AND DISCUSSION}
 First, we discuss the overall band structure of MnTe in the AF phase. We carried out $h\nu$-dependent ARPES measurements and estimated the $k_z$ value (see Fig. 6 of Appendix\ \hyperref[apxB]{B}). The contour map of ARPES intensity at $E_\textrm{F}$ for the AHL plane obtained with $h\nu=117 \textrm{ eV}$ in Fig. 1(c) signifies a bright spot centered at each A point that follows the periodicity of hexagonal BZ. The band dispersion along the AH high-symmetry cut shown in Fig. 1(d) signifies a dispersive hole band (white arrow) at the A point, together with a shallower hole band outside this band (red arrow). These bands produce a spot centered at the A point in Fig. 1(c). One can also recognize in Fig. 1(d) a broad feature at the binding energy ($E_\textrm{B}$) of $\sim$2 eV around the A point (yellow arrow) and another broad feature which weakly disperses upward on approaching the H point (blue arrow). We carried out the ARPES measurements also for the $\Gamma\textrm{KM}$ plane with $h\nu=80 \textrm{ eV}$ and found that the band dispersion [Fig. 1(e)] and the energy contour [Fig. 1(f)] are markedly different. As shown in Fig. 1(e), instead of a simple hole band, the top of the valence band along the $\Gamma\textrm{K}$ cut shows an M-shaped dispersion with its maxima slightly away from the $\Gamma$ point (green arrow) producing six petals surrounding the $\Gamma$ point in Fig. 1(f). 
 
 To discuss the band character in more detail, 
we show in Fig. 1(g) the energy distribution curves (EDCs) around $E_\textrm{F}$ along the $\Gamma\textrm{K}$ cut. The result signifies the band splitting with a maximum value of $\sim$0.3 eV at slightly away from the $\Gamma$ point. This band splitting is also seen in Fig. 1(e) (green and purple arrows) and associated with SOC. 
We found that the DFT calculations with a spin configuration and experimental geometry of either ($S_{a{\parallel}}, \Gamma\textrm{K}_2$) or ($S_{a{\perp}}, \Gamma\textrm{K}_1$) setting reproduce well the experimental results (for details, see  Appendix\ \hyperref[apxC]{C}). Since these two cases show almost identical band structure, we hereafter adopt the former case just for our convenience.
 
 As shown in Fig. 1(h), the experimental band structure along the $\textrm{AH}$  cut is well reproduced by the band calculation. In particular, the steeper inner hole band (white arrow) and the outer hole band (red arrow), as well as the feature at  $E_\textrm{B} \sim$ 1.3--2.0 eV (yellow and blue arrows) seen by ARPES are well reproduced by the calculation. These bands are attributed to the hybridized Te-5$p$ and Mn-3$d$ $t_{2g}$ orbitals. We also found an overall agreement of the band dispersion between the experiment and calculation along the $\Gamma\textrm{K}$ cut [Fig. 1(i)]. The valence-band top is experimentally located at the A point around $E_\textrm{F}$, in line with the calculation, signifying the semiconducting nature of MnTe. Taking into account the total band-gap size of $\sim$1.3 eV \cite{Allen1977, FerrerPRB2000}, it is suggested that MnTe is a $p$-type semiconductor, consistent with the transport measurement \cite{Wasscher1969, KriegnerNC2016}. 
 
 Next we focus on the observation of the altermagnetic band-splitting. Since the altermagnetic band-splitting was predicted to have nodes along high-symmetry lines under negligible SOC, it is necessary to carry out ARPES measurements along non-high-symmetry cuts to observe the altermagnetic band-splitting. Non-high-symmetry $\Gamma\textrm{L}$ cut [see Fig. 2(a)] is suited for this sake because the DFT calculations identified a sizable splitting along this cut \cite{SmejkalPRX2022_1, Rooj2024}. On the other hand, the ARPES measurement needs a special care because $h\nu$ must be changed at every step of momentum to correctly sweep $\textbf{k}$ along the $\Gamma\textrm{L}$ cut. Since the altermagnetic band-splitting can show up even without SOC, we first explain the calculated band structure without SOC shown in Fig. 2(b). 
 One can recognize the spin-split bands in a whole ($E, \textbf{k}$) region. Although many bands are observed in this region, one can easily identify a spin-split partner of each band, thanks to the band degeneracy along high-symmetry lines. For example, bands labelled 1 and 2 (sequentially labeled from the highest occupied band around $\Gamma$) are spin-split partners to each other because they degenerate at the  $\Gamma$ and L points. The altermagnetic band-splitting associated with these bands are strongly anisotropic; it reaches $\sim$0.8 eV at around $k_z \sim 0.5\pi$, whereas it is zero at the $\Gamma$ ($k_z =0$) and L ($k_z = \pi$) points. Similarly, the altermagnetic band-splitting for bands 3 and 4 has nodes at the $\Gamma$ and L points and takes a maximum at $\sim$1.0 eV around $k_z \sim 0.7\pi$. The band 2 intersects the band 3 due to the large spin splitting. Such a large splitting is associated with the  $\mathcal{PT}$ (space- and time-reversal) symmetry breaking of the Te site and strong hybridization between Te-5$p$ and Mn-3$d$ orbitals. When the SOC is included in the calculation, these bands hybridize each other to produce a small spin-orbit gap at the intersection, but the overall spin-splitting feature is essentially preserved [Fig. 2(c)]. Although the inclusion of SOC makes the assignment of spin-spilt partners not so straightforward, one can discuss the characteristics of the spin splitting by referring to the calculations without SOC. As shown in Fig. 2(d), the experimental band structure along the $\Gamma\textrm{L}$ cut shows a rough agreement with the calculation [Fig. 2(b) or 2(c)].
  
 To investigate the altermagnetic band-splitting in more detail, we performed ARPES measurements along representative $\textbf{k}$ cuts which cross the $\textbf{k}$ points in the $\Gamma\textrm{KHL}$ plane including the $\Gamma\textrm{L}$ cut [cuts 1--5 in Fig. 2(a)]. Cut 1 with ($k_x, k_z) = (0, 0)$ and cut 5 with $(k_x, k_z) = (\pi, \pi)$ correspond to the  $\Gamma\textrm{K}$ and LH high-symmetry cuts, respectively, whereas cuts 2--4 trace non-high-symmetry $\textbf{k}$ points. Figures 2(g), 2(e), and 2(f) show the obtained ARPES intensity, the corresponding calculated band structures without SOC, and those with SOC, respectively. The calculated bands along the $\Gamma\textrm{K}$ cut (cut 1) in Fig. 2(e1) are spin degenerate irrespective of $\textbf{k}$ when the SOC is neglected, as highlighted by the case of bands 1 and 2 (thick black curves). When the SOC is included, the degeneracy is lifted in most of the $\textbf{k}$ points except for the $\Gamma$ point. The spin polarization remains zero along the  entire $\Gamma\textrm{K}$ line due to the two-fold rotation and mirror symmetries of the crystal [Fig. 2(f1)]. As the $\textbf{k}$ cut moves away from the high-symmetry $\textbf{k}$ points, the bands start to spin split [cut 2; Fig. 2(e2)]. In particular, bands 1 and 2 indicated by thick red and blue curves, respectively, show a strongly $\textbf{k}$-dependent exchange splitting accompanied by the sign reversal upon varying $k_y$, in contrast to the case of a simple itinerant ferromagnet showing a $\textbf{k}$-independent exchange splitting. We found that the $\textbf{k}$ points in which the sign reversal occurs are exactly on the $\overline{\Gamma\textrm{K}}$ high-symmetry lines of the surface BZ. A similar anisotropic band-splitting is also recognized along cuts 3 and 4 [Figs. 2(e3) and 2(e4)] whereas the overall magnitude is gradually suppressed and eventually vanished along the high-symmetry LH cut [cut 5; Fig. 2(e5)]. Intriguingly, the magnitude of calculated altermagnetic band-splitting reaches +0.8 eV at $|k_y| \sim 0.25 \pi$ and -0.8 eV at $|k_y| \sim 0.5 \pi$ along cut 3 (we define a positive splitting when the energy level of band 1 is higher than that of band 2, and {\it vice versa} for the negative splitting).
 
 The experimental band structure [Fig. 2(g)] shows a reasonable agreement with the calculated band dispersion including SOC [Fig. 2(f)]. For example, one can recognize a common trend that the topmost band gradually sinks downward and becomes flatter on moving from cut 1 to cut 5 in both the experiment and calculation. Along cut 1 ($\Gamma\textrm{K}$  cut), the SOC-induced band splitting of the topmost valence band (green and purple arrows) is reproduced by the calculation as already shown in Fig. 1. Along cut 3, the calculation signifies two local maxima in the band dispersion at the $\Gamma$  point for band 1 (red arrow) and slightly away from the $\Gamma$ point for band 2 ($|k_y| \sim 0.4\pi$; blue arrow). Since these bands are the spin-split partners, the experimental observation of a double-peaked dispersion along cut 3 supports the existence of altermagnetic band-splitting.  We also find a signature of the band bottom (yellow arrow) which may correspond to the local minimum of calculated band 1. To trace the energy dispersion of split bands in more straightforward way, we show in Fig. 2(h) a series of EDCs obtained along cut 3. One can recognize peaks corresponding to band 1 (red color) and band 2 (blue color) which seem to intersect at specific $k_y$ positions (e.g., at $|k_y| \sim 0.21\pi$ and $0.71\pi$), consistent with the calculated band structure shown in Fig. 2(e3) (strictly speaking, these bands do not intersect due to the spin-orbit gap, whereas the gap is experimentally unclear probably due to the lifetime broadening and/or $k_z$ broadening). Importantly, at $|k_y| = 0.43 \pi$, the bands 1 and 2 are located at the binding energy of 1.25 and 0.45 eV, respectively, corresponding to the altermagnetic band splitting of $\sim$0.8 eV, which has a magnitude comparable to that of the calculation. Moreover, one can see that, not only the general trend of band dispersion, but also the overall energy position of bands 1 and 2 is similar between the experiment and calculation. In fact, the yellow and blue arrows indicated in the second-derivative plot in Fig. 2(g3) well coincide with the peak positions of the EDCs for the bands 1 and 2 in Fig. 2(h). These support our assignment of the experimental band dispersion, namely, the altermagnetic band-splitting between these bands. We found that the overall agreement between the experiment and calculation becomes rather poor along cut 4. This is likely due to the smaller band splitting that causes a difficulty in experimentally distinguishing the bands 1 and 2, consistent with the general trend that the calculated band splitting takes the largest value at  $k_z \sim0.4\pi$ (cut 3). It is noted that the bright feature at $E_\textrm{B}$ = 1.3--2.0 eV in Fig. 2(g4) is likely associated with band 3, while the energy position is slightly lower than the calculation probably due to the $k_z$ broadening effect in the experiment and/or the overestimation of band energy in the calculation. It is also remarked that the band width in the experiment in Figs. 2(g1)--(g5) appears to be slightly different from that in the calculation in Figs. 2(f1)--(f5), as inferred from the narrower band width in the experiment along cut 1 [compare energy position of bands at the K point between Figs. 2(f1) and 2(g1)]. Such a difference may be associated with the electron correlation effect which is not taken into account in the calculation.

To further validate the altermagnetic origin of the observed band splitting, we have performed temperature-dependent ARPES measurements at $|k_y|=0.5\pi$ along cut 3 where the altermagnetic band-splitting is clearly observed [highlighted by a thick EDC in Fig. 2(h)]. As shown in Fig. 3(a), at $T$ = 30 K, one can recognize two peaks at  $E_\textrm{B}$ = 1.2 and 0.6 eV due to the altermagnetic band-splitting. On increasing temperature, two peaks gradually merge and eventually become indistinguishable at $T \sim 300\text{--}310 \textrm{ K}$. 
This systematic evolution can hardly be explained in terms of a simple change in the spectral weight, as suggested from our numerical simulation (for details, see  Appendix\ \hyperref[apxD]{D}). The temperature at which the two peaks merge coincides with $T_\mathrm{N}$, suggesting that the splitting is associated with the AF transition. Such a change in the band splitting is also supported by numerical simulations of the EDC at each temperature that take into account the single/double Voigt-function peak(s) and moderate background, as highlighted in Figs. 3(b) and 3(c) where the EDCs at $T=310 \textrm{ K}$ (above  $T_\mathrm{N}$) and $60 \textrm{ K}$ (well below $T_\mathrm{N}$) are reasobly reproduced by the single and double peaks, respectively. All these results provide a spectroscopic evidence for the existence of altermagnetic band-splitting in the colinear AF phase of MnTe. A next challenge is to accurately determine the $\textbf{k}$-dependent spin texture by spin-resolved micro-ARPES measurement.
 
 To highlight our key findings, we draw in Fig. 4(a) the schematic band dispersion for the spin-split partners, bands 1 and 2, in the $\Gamma\textrm{KHL}$ plane based on the DFT calculation supported by the ARPES observation. When the SOC is neglected, these bands are degenerate along the $\Gamma\textrm{K}$ and LH high-symmetry lines, producing dispersive nodal lines (orange curves) across which the sign of altermagnetic band-splitting is reversed. There exist additional nodal lines running along the $\overline{\Gamma\textrm{K}}$ cut. All these nodal lines are regarded as a slice of nodal surfaces in the $\Gamma\textrm{KHA}$ and MKHL planes. The existence of multiple nodal lines is also visualized by the intensity plot of calculated altermagnetic band-splitting in hexagonal BZ in Fig. 4(b) in which one can recognize the nodal lines running along the $\overline{\Gamma\textrm{K}}$ and $\overline{\textrm{KM}}$ high-symmetry lines of the surface BZ as well as the sign reversal of altermagnetic band-splitting across the nodal lines. Such sign-reversal band splitting is a unique characteristic of altermagnets, distinct from conventional sign-preserving ferromagnetic exchange splitting.
 
 When the SOC is included, the band degeneracy of the nodal surfaces is lifted by the spin-orbit gap in entire BZ, whereas the bands are still spin degenerate along some particular high-symmetry lines such as the $\Gamma\textrm{A}$ line, protected by the $C_{2z}$ symmetry of the crystal. Low-energy excitations of such nodal line are characterized by the altermagnetic quasiparticle with a quadratic band dispersion \cite{SmejkalPRX2022_1}. Importantly, since the overall energy scale of the spin-orbit gap ($< 0.3 \textrm{ eV}$) is smaller than that of the altermagnetic band-splitting ($< 0.8 \textrm{ eV}$), the anisotropic sign-reversal exchange splitting, a key characteristic of altermagnets, is still maintained even under the presence of SOC. The present result thus establishes that the theoretically predicted altermagnetic band-splitting is indeed realized in MnTe. It is emphasized that the altermagnetic band-splitting verified in this study would be responsible for the piezomagnetic effect and the x-ray magnetic circular dichroism recently reported for MnTe \cite{AoyamaArXiv2023, HarikiArXiv2023}, both of which are sensitive to the global TRS breaking. Also, the multiple band crossing and sign reversal of the altermagnetic band-splitting inherent to MnTe would give rise to a high Berry curvature region in $\textbf{k}$ space and may be responsible for the spontaneous anomalous Hall effect as experimentally verified recently \cite{GonzalezPRL2023}.
 
 The micro-ARPES measurements and the DFT calculations have established the presence of altermagnetic band-splitting in the colinear AF phase of MnTe. The altermagnetic band-splitting is strongly $\textbf{k}$  dependent and reaches $\sim$0.8 eV at non-high-symmetry  $\textbf{k}$ points in the $\Gamma\textrm{KHL}$ plane, whereas the splitting at the high-symmetry $\textbf{k}$ points is suppressed, producing multiple nodal lines. We also observed a small ($\sim$0.3 eV) band splitting along the high-symmetry $\Gamma\textrm{K}$ cut, suggestive of an additional contribution from the SOC. The present results lay a foundation for exploring unique physical properties inherent to the altemagnetic MnTe, paving a pathway toward investigating the altermagnetic band-splitting in other altermagnets.
 
\begin{acknowledgments}
We thank K. Ozawa, M. Kitamura, K. Horiba, and H. Kumigashira for their assistance in the ARPES experiments. This work was supported by JST-CREST (No. JPMJCR18T1 and JP19198318), JST-PRESTO (No. JPMJPR18L7), and Grant-in-Aid for Scientific Research (JSPS KAKENHI Grant Numbers JP21H04435, JP19H01845, JP22H00102, JP19H05823, and JP19H05822), Grant-in-Aid for JSPS Research Fellow (No: JP18J20058), and KEK-PF (Proposal number: 2021S2-001). A.H. thanks GP-Spin and JSPS for financial support.
\end{acknowledgments}

\clearpage

\appendix
\section{SAMPLE CHARACTERIZATION}
\label{apxA}We show in Fig. 5(a) a photograph of a typical single crystal used in this study. Details of the sample preparation and characterization were described elsewhere \cite{AoyamaArXiv2023}. Orientation of the crystal and a good single crystallinity were confirmed by x-ray Laue backscattering measurement which signifies clear six-fold symmetric diffraction spots as seen in Fig. 5(b). The clean surface nature of the crystal cleaved in ultrahigh vacuum of  $1\times10^{-10} \textrm{ Torr}$ was confirmed by sharp core-level peaks originating from the Mn 3$p$ and Te 4$d$ orbitals with no inclusion of contaminant peaks, as shown in Fig. 5(c).

\section{NORMAL-EMISSION  ARPES SPECTRA}
\label{apxB}Figures 6(a) and 6(b) show the EDCs and corresponding ARPES intensity, respectively, along the wave vector perpendicular to the sample surface ($k_z$), measured with the normal-emission setup by varying $h\nu$ in the VUV region ($h\nu=60\text{--}170 \textrm{ eV}$). One can find some energy bands displaying a finite $k_z$ dispersion, e.g., in $E_\textrm{B}$ range of $E_\textrm{F}$--0.7, and 1.2--2.6 eV. The observed band dispersions well follow the periodicity of the bulk Brillouin zone as well as the calculated band dispersion including SOC (green curves). The periodic dispersion along the $k_z$ direction indicate the bulk origin of the observed bands.

\section{BAND-STRUCTURE CALCULATIONS WITH DIFFERENT IN-PLANE SPIN CONFIGURATION}
\label{apxC}In the presence of SOC, the energy eigenvalue in AF phase generally differs depending on spin direction \cite{PauloPRB2023}. Although the magnetic torque and neutron diffraction experiments \cite{Komatsubara1963, Szuszkiewicz2005, KriegnerPRB2017} have suggested the in-plane nature of the magnetic moment in the AF phase, it is experimentally unclear whether the magnetic moment is directed along the $a$-axis corresponding to the nearest neighbor Mn-Mn bonding direction ($k_y$ axis in the $k$ space) or perpendicular to it ($k_x$ axis in $k$ space). To examine to what extent the choice of possible spin configurations and selection of inequivalent $\textbf{k}$ cuts influence the calculated band structure, we carried out DFT calculations for bulk MnTe with spin configurations parallel and perpendicular to the $a$ axis, called here, $S_{a\parallel}$ and $S_{a\perp}$, respectively. Here we focus on the band dispersion along the $\Gamma\textrm{K}$ cuts, because the band splitting of the topmost valence band is useful to pin down a possible spin configuration and actual $\textbf{k}$ cuts in the experiment. For the $S_{a\parallel}$ configuration, there exist two inequivalent $\Gamma\textrm{K}$ cuts, one directed along the spin axis, $\Gamma\textrm{K}_1$, and the other rotated by $60^{\circ}$ from it, $\Gamma\textrm{K}_2$ [Fig. 7(b)]. For the calculation with $S_{a\perp}$ configuration, we fix these $\textbf{k}$ cuts but just rotate the spin direction by $90^{\circ}$. As shown in Fig. 7(c) for the $S_{a\parallel}$ configuration, the band dispersion for the topmost valence band shows no band splitting along the $\Gamma\textrm{K}_1$ cut, whereas it apparently splits into two bands (bands 1 and 2) along the $\Gamma\textrm{K}_2$ cut [Fig. 7(d)] with a maximum splitting size of 0.25 eV. Such a difference is due to the symmetry difference of these two $\textbf{k}$ cuts with respect to the magnetic moment. For the $S_{a\perp}$ configuration, the $\Gamma\textrm{K}_1$ cut exhibits a band splitting comparable to that of the  $\Gamma\textrm{K}_2$ cut for the $S_{a\parallel}$ configuration, whereas the band splitting along the $\Gamma\textrm{K}_2$ cut is overall reduced.
We show in Fig. 7(k) the experimental band dispersion determined by tracing the numerical fittings to the EDCs in Fig. 1(g), in comparison with the calculated band structure for four sets of possible spin configurations and $\textbf{k}$ cuts. Taking into account that the experimental band splitting is as large as $\sim$0.3 eV along the $\Gamma\textrm{K}$ cut, it is likely that the experimental spin and measurement geometry is either ($S_{a\parallel}$, $\Gamma\textrm{K}_2$) or ($S_{a\perp}$, $\Gamma\textrm{K}_1$). As shown in Figs. 7(g)-7(j), there are no much differences in the band dispersion along cut 3. This suggests that the key band argument based on cut 3 in the main text is not much affected by the choice of $\textbf{k}$ cuts and spin configurations.

\section{TEMPERATURE-DEPENDENT ARPES SPECTRA}
\label{apxD}To examine in more detail temperature-dependent change in the EDCs shown in Fig. 3(a), we show in Fig. 8(a) a plot of the EDCs with no vertical offset. One can see that energy position of the first peak located at closer to $E_\textrm{F}$ apparently moves toward higher binding energy upon increasing temperature. This systematic shift can hardly be explained in terms of the simple change in the spectral weight. To further examine this point, we have performed numerical simulations of the EDC at $T = 310 \textrm{ K}$ by assuming two peaks with their energy positions fixed to those at $T = 30 \textrm{ K}$. As a result, we found that it is difficult to reproduce the experimental EDC, no matter what broadening and spectral-weight parameters we use. This is highlighted by a simulation curve shown by a brown curve (best fit) in Fig. 8(b) in which one can recognize poorer agreement with the experimental EDC (open circles) compared to the numerical simulation with a single peak [green curve; same as Fig. 3(b)]. These results suggest intrinsic reduction of the exchange splitting upon increasing temperature.

\bibliographystyle{apsrev4-2ss}

\bibliography{MnTe_refs}

\clearpage
\begin{figure}
\begin{center}
\includegraphics[width=4in]{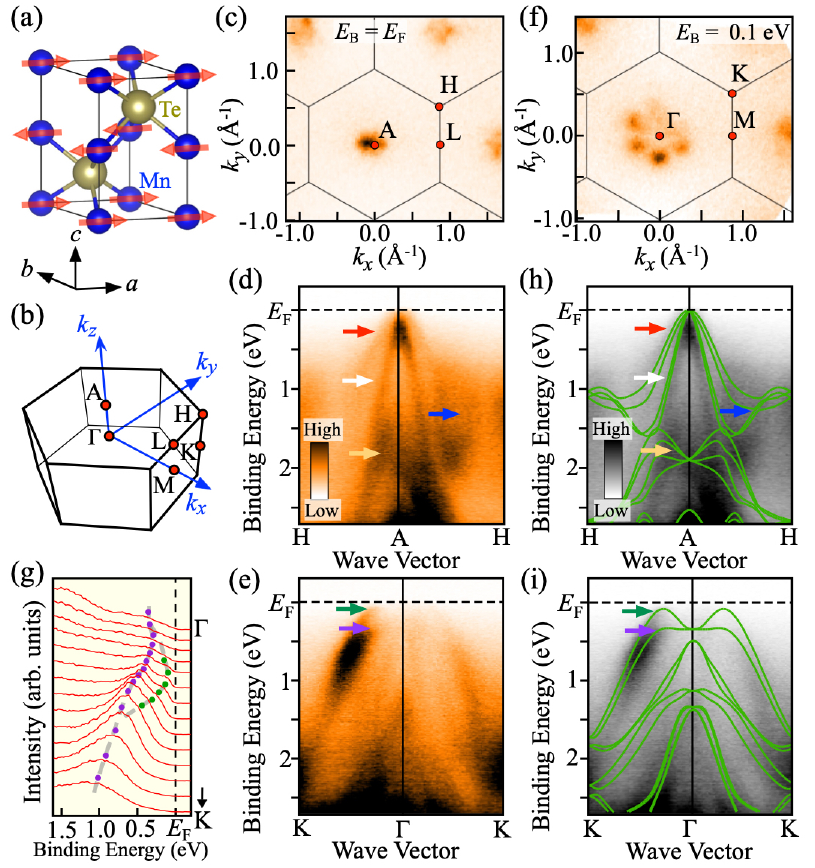}
\hspace{0.2in}
\caption{(a),(b) Crystal structure and bulk hexagonal BZ of MnTe, respectively. Spin configuration for the $S_{a{\parallel}}$ case is indicated by red arrows in (a). (c) ARPES intensity maps at $T=40 \textrm{ K}$ as a function of $k_x$ and $k_y$ at $E_\textrm{B} = E_\textrm{F}$ for $k_z\sim\pi$ ($\textrm{AHL}$) plane. (d),(e) ARPES intensity as a function of wave vector and $E_\textrm{B}$ measured along the $\textrm{AH}$ cut of bulk BZ obtained at $h\nu=117 \textrm{ eV}$ and the $\Gamma\textrm{K}$ cut at $h\nu=80 \textrm{ eV}$, respectively. (f) Same as (c), but at $E_\textrm{B} = 0.1 \textrm{ eV}$ for the $k_z \sim 0$ ($\Gamma\textrm{KM}$) plane. (g) EDCs near $E_\textrm{F}$ along the  $\Gamma\textrm{K}$ cut. (h),(i) Calculated band structures along the AH and $\Gamma\textrm{K}$ cuts, respectively, for the AF phase with the ($S_{a{\parallel}}, \Gamma\textrm{K}_2$) configuration with including the spin-orbit coupling (SOC), overlaid with the ARPES intensity with gray scale.
}
\end{center}
\end{figure}

\clearpage
\begin{figure*}
\begin{center}
\includegraphics[width=6.5in]{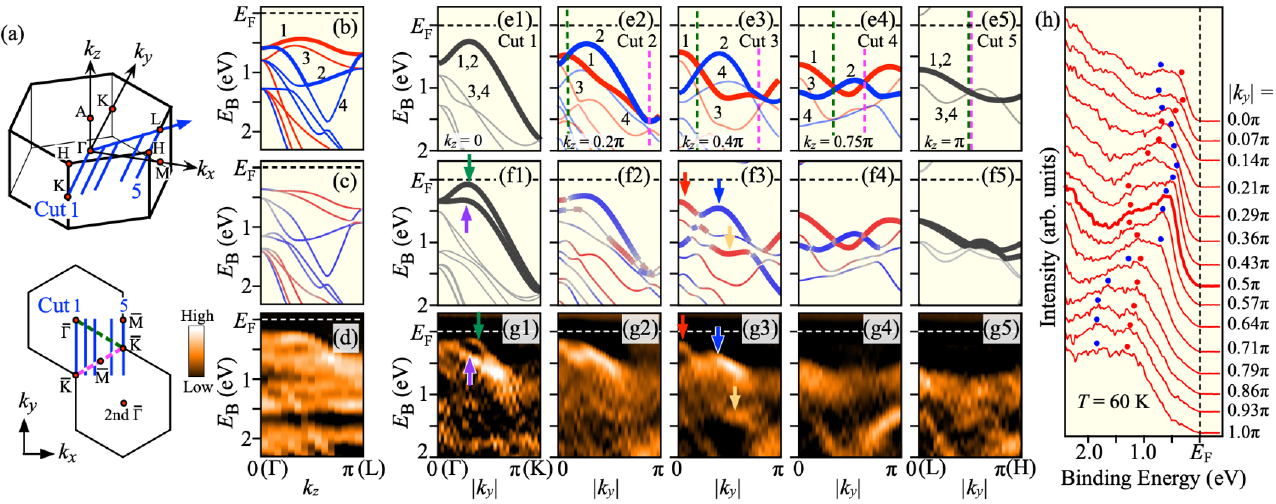}
\hspace{0.2in}
\caption{(a) Bulk BZ (top) and surface BZ (bottom) of MnTe, together with the $k$ cuts (blue solid lines) where calculations and ARPES measurements shown in (e)--(g) were carried out. (b),(c) Calculated band structure along the $\Gamma\textrm{L}$ cut in the AF phase without and with SOC, respectively. Red and blue curves correspond to up and down spin components respectively, in which the quantization axis for spins is defined along the $k_y$ axis. Gray color corresponds to the zero spin polarization. (d) Second-derivative plot of ARPES intensity at $T=40 \textrm{ K}$ along the $\Gamma\textrm{L}$ cut. (e),(f) Calculated band structures in the AF phase without and with SOC, respectively, along the $k_y$ cut passing through the  $\Gamma\textrm{L}$ line (cuts 1-5) shown by blue lines in (a). Thick red and blue curves highlight bands 1 and 2, respectively. Calculations for cuts 1--5 were carried out at $(k_x, k_z) = (0, 0), (0.2\pi, 0.2\pi), (0.4\pi, 0.4\pi), (0.75\pi, 0.75\pi), \textrm{and } (\pi, \pi)$, respectively. $|k_y| = \pi$ corresponds to the $\Gamma\textrm{K}$ length.  Green and purple dashed lines correspond to the $k$ point on the $\overline{\Gamma\textrm{K}}$ high-symmetry line of the surface BZ, as also indicated in the bottom panel of (a). (g) Corresponding second-derivative plots of ARPES intensity as a function of $|k_y|$ measured along cuts 1--5. (h) EDCs corresponding to the band mapping of (g3) at $T=60 \textrm{ K}$. Red and blue dots trace the bands 1 and 2, respectively. Yellow and blue arrows shown in (g3) are also plotted.  
}
\end{center}
\end{figure*}

\clearpage
\begin{figure}
\begin{center}
\includegraphics[width=4in]{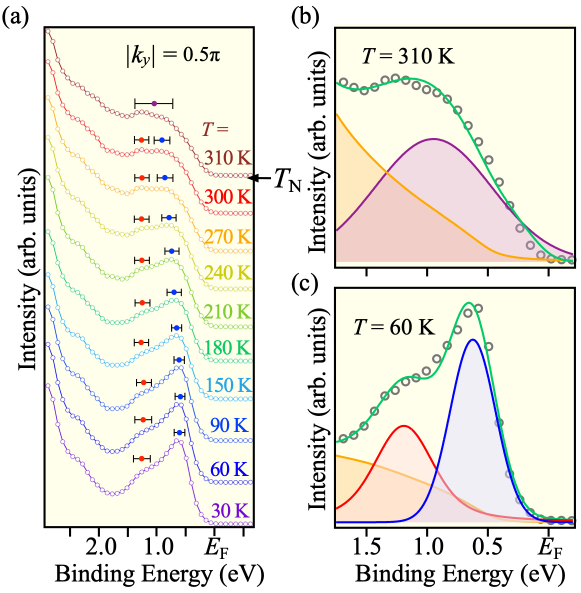}
\hspace{0.2in}
\caption{(a) Temperature dependence of EDC at $|k_y| = 0.5\pi$ shown by a thick curve in Fig. 2(h). Red/blue dots show the energy position of peaks estimated by numeric simulations of EDCs assuming Voigt-function peak(s) and moderate background. (b),(c) Results of numerical simulations at $T=310 \textrm{ K}$ (above $T_\textrm{N}$) and 60 K (below $T_\textrm{N}$), respectively.
}
\end{center}
\end{figure}

\clearpage
\begin{figure}
\begin{center}
\includegraphics[width=5.2in]{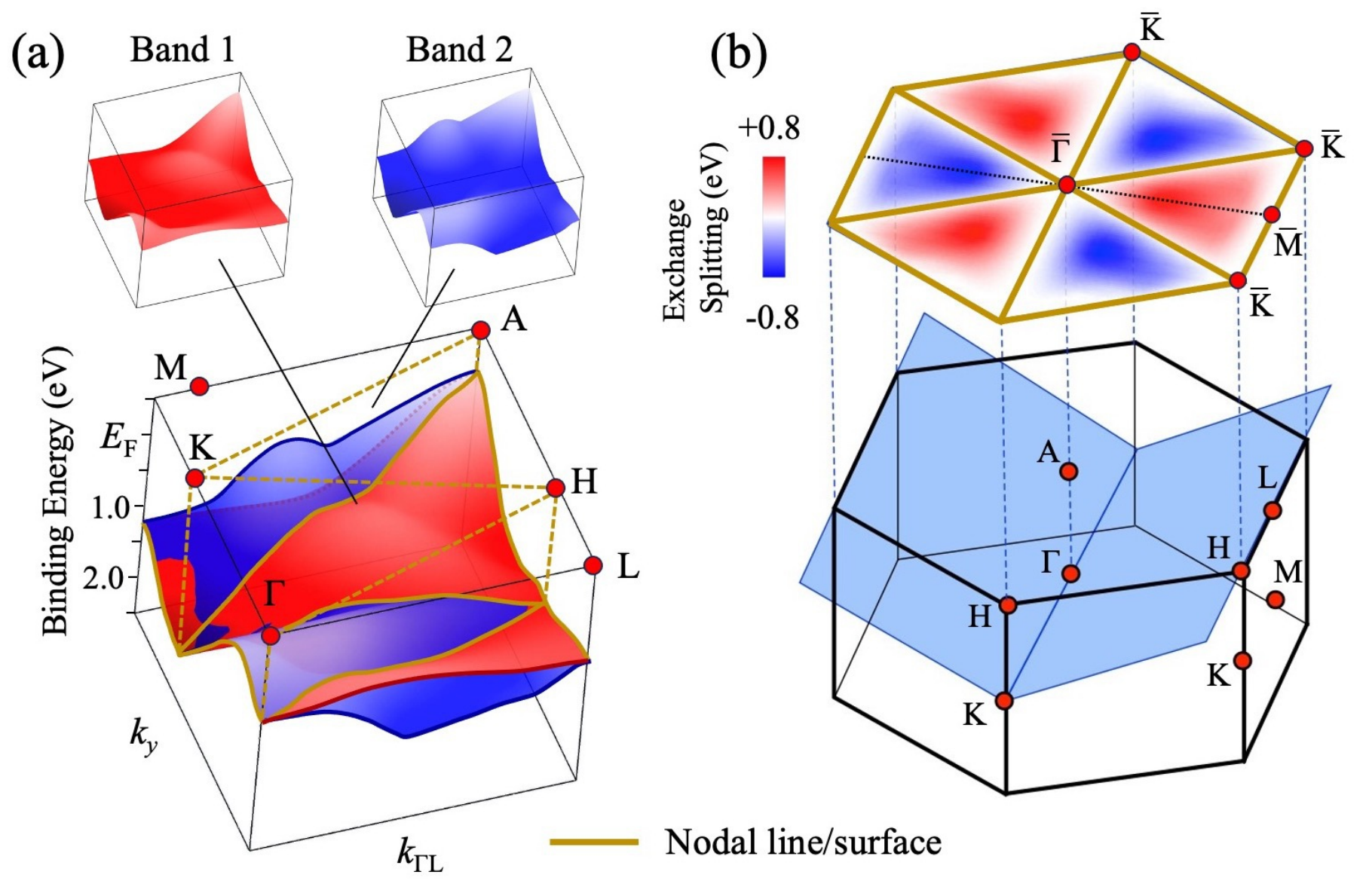}
\hspace{0.2in}
\caption{(a) Schematic 3D band dispersion of MnTe in the $\Gamma\textrm{KHL}$ plane without SOC, signifying the dispersive Dirac-nodal lines running along the $\overline{\Gamma\textrm{KM}}$ high-symmetry lines of the surface BZ (orange curves). (b) $\Gamma\textrm{KHL}$ plane (shaded area) in 3D BZ in which the band dispersion in (a) is shown. Intensity plot of calculated altermagnetic band-splitting in the hexagonal surface BZ is shown in the upper panel.
}
\end{center}
\end{figure}

\clearpage
\begin{figure}
\begin{center}
\includegraphics[width=5.2in]{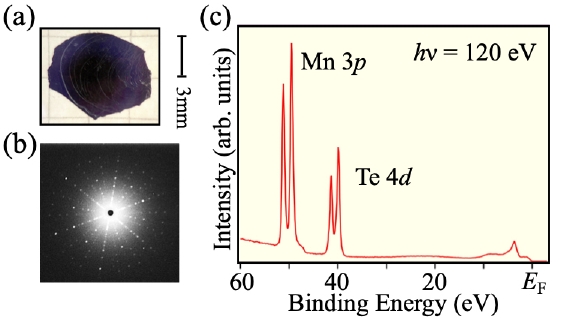}
\hspace{0.2in}
\caption{(a) Photograph of a MnTe single crystal. (b) Representative Laue x-ray diffraction pattern of the (0001) surface. (c) Photoemission spectrum of MnTe in a wide energy region obtained at  $h\nu=120 \textrm{ eV}$.
}
\end{center}
\end{figure}

\clearpage
\begin{figure}
\begin{center}
\includegraphics[width=5.2in]{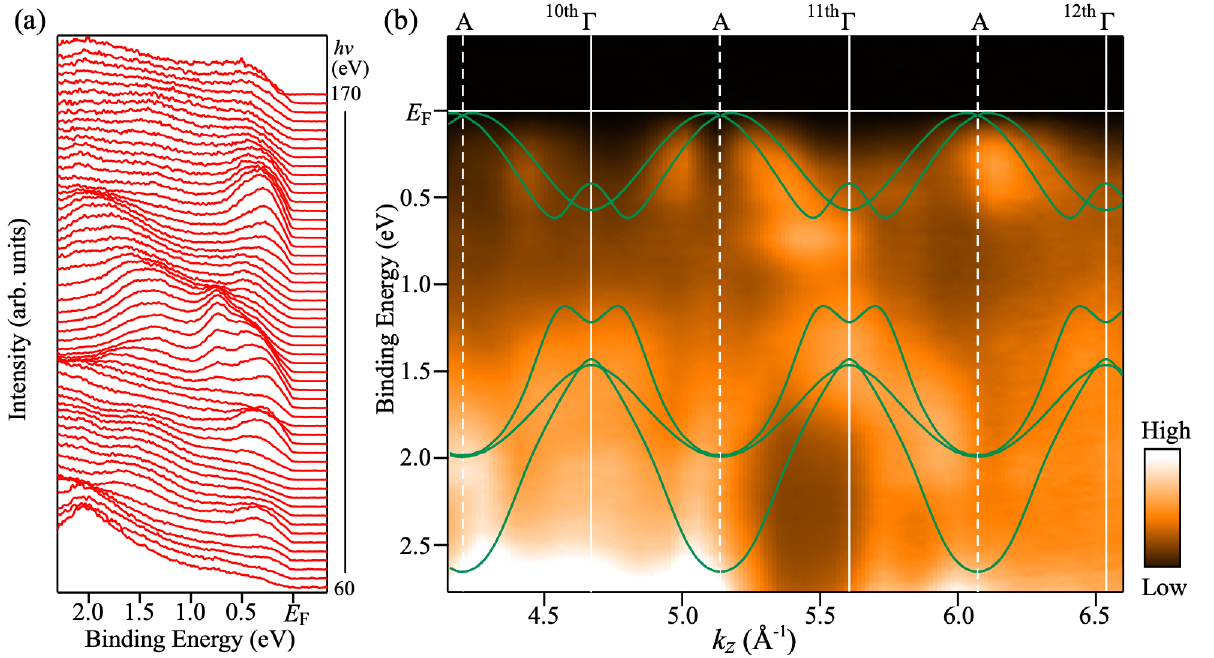}
\hspace{0.2in}
\caption{(a),(b) Plots of the normal-emission EDCs and corresponding ARPES intensity, respectively, as a function of $h\nu$ or $k_z$ (corresponding to the out-of-plane $\Gamma\textrm{A}$ cut). Inner potential was set to be $V_0=8.0 \textrm{ eV}$ from the periodicity of the band dispersion. Green curves represent the calculated band structure including SOC along the $\Gamma\textrm{A}$ cut, for the ($S_{a\parallel}$, $\Gamma\textrm{K}_2$) setting.
}
\end{center}
\end{figure}

\clearpage
\begin{figure*}
\begin{center}
\includegraphics[width=6.5in]{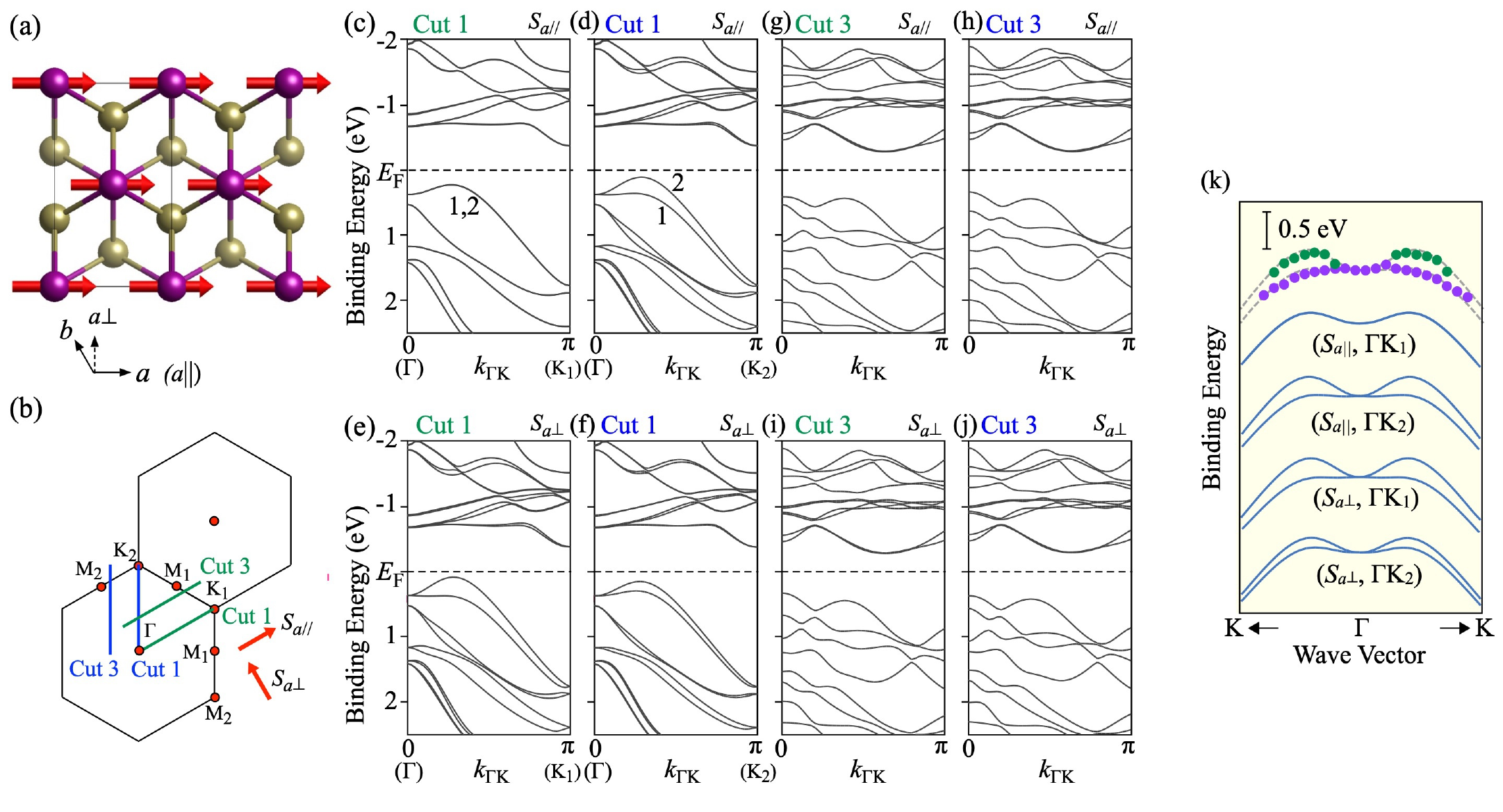}
\hspace{0.2in}
\caption{(a) Crystal structure with in-plane spin configuration, called here $S_{a\parallel}$. (b) $k$ cuts in surface BZ along which DFT calculations were carried out (green and blue lines). Direction of spin vectors, $S_{a\parallel}$ and $S_{a\perp}$, is also indicated. We specify two inequivalent K points as $\textrm{K}_1$ and $\textrm{K}_2$, and also for the M points ($\textrm{M}_1$ and $\textrm{M}_2$). As a result, the $\Gamma\textrm{K}$ cut [cut 1; $k_z$ = 0] has two configurations, namely, $\Gamma\textrm{K}_1$ and $\Gamma\textrm{K}_2$. (c),(d) Calculated band structure with SOC for the $S_{a\parallel}$ configuration along $\Gamma\textrm{K}_1$ and $\Gamma\textrm{K}_2$ cuts, respectively. (e),(f), Same as (c) and (d) but for the $S_{a\perp}$ configuration. (g)--(j) Same as (c)--(f) but for cut 3 ($k_z = 0.4\pi$). Definition of cuts 1 and 3 is the same as that in the main text. (k) Experimental band dispersion obtained by fitting to the EDCs in Fig. 1(g), compared with the calculated band structure along the $\Gamma\textrm{K}$ cut in the AF phase for the four sets of spin configurations.
}
\end{center}
\end{figure*}

\clearpage
\begin{figure}
\begin{center}
\includegraphics[width=4in]{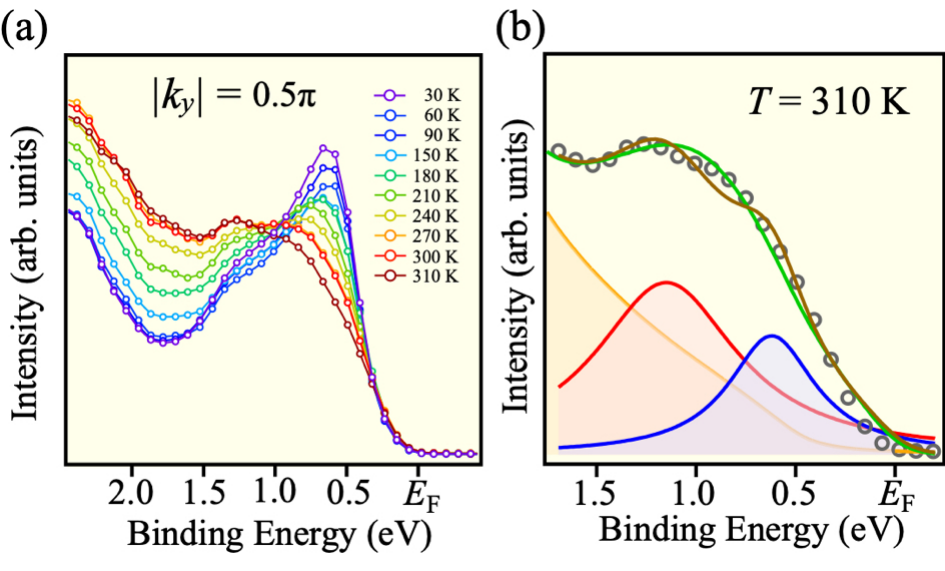}
\hspace{0.2in}
\caption{(a) Same as Fig. 3(a) but without a vertical offset. (b) Result of numerical simulation of the EDC at $T = 310 \textrm{ K}$ by assuming two peaks with their peak positions fixed to those at $T = 30 \textrm{ K}$ (brown curve). Green curve represents numerical simulation by assuming a single peak [same as Fig. 3(b)].
}
\end{center}
\end{figure}

\end{document}